\documentclass[12pt]{iopart}

\newcommand{\be}{\begin{equation}}
\newcommand{\ee}{\end{equation}}
\newcommand{\bea}{\begin{eqnarray}}
\newcommand{\eea}{\end{eqnarray}}
\newcommand{\iGA}{{i}}

\usepackage{iopams}  

\usepackage{graphicx}

\begin{document}

\title[Multivector Spacetime]{An explanation for galaxy rotation curves using a Clifford multivector spacetime framework}

\author{James M.~Chappell, Nicolangelo Iannella, Azhar Iqbal and Derek Abbott}

\address{School of Electrical and Electronic Engineering, University of Adelaide, South Australia 5005, Australia}

\ead{james.m.chappell@adelaide.edu.au}

\begin{abstract}
We explore the consequences of space and time described within the Clifford multivector of three dimensions $ Cl_{3,0}$, where space consists of three-vectors and time is described with the three bivectors of this space.  When describing the curvature around massive bodies, we show that this model of spacetime when including the Hubble expansion naturally produces the correct galaxy rotation curves without the need for dark matter.
\end{abstract}


\maketitle

\section{Introduction}
The concept of time is typically modeled as a fourth Euclidean-type dimension appended to the three dimensions of spatial movement forming Minkowski spacetime \cite{SexlUrbantke2001}.  However with the observed non-Cartesian like behavior of time, such as the time axis possessing a negative contribution to the Pythagorean distance, and the observed inability to freely move within the time dimension, as is possible with space dimensions, suggests that an alternate representation might be preferable.

Recently a Clifford multivector algebraic model for spacetime was developed \cite{chappell2011revisiting}, with time represented as a bivector.  A bivector in geometric algebra represents an oriented unit area, which also acts as a rotation operator.  Therefore in three-dimensions we have a three-vector $ \mathbf{x} $ and time becomes a bivector oriented within three dimensions, represented as $ i c \mathbf{t} $ where $ \mathbf{t} $ is also a three-vector.  These can then be added to form a spacetime multivector $ \mathbf{x} + i  c \mathbf{t} $. This now gives rise to a natural symmetry between the space and time coordinates, both being represented as three-vectors, while still recovering the standard results of special relativity.  

Clifford's geometric algebra was first published in 1873, extending the work of Grassman and Hamilton, creating a single unified real mathematical framework over Cartesian coordinates, which included the algebraic properties of scalars, complex numbers, quaternions and vectors into a single entity, called the multivector \cite{Doran2003}.  We find that this general algebraic entity, as part of a real three-dimensional Clifford algebra $ { \it Cl}_{3,0}(\Re) $, provides an equivalent representation to a Minkowski vector space $ \Re^{3,1} $ \cite{rodrigues2007many,matolcsi1984models} as well as a suitable description of curved spacetime used in general relativity.
 
In order to represent the three independent degrees of freedom of space, Clifford defined three algebraic elements $ e_1 $, $ e_2 $ and $ e_3 $, with the properties
\be \label{orthonormality}
e_1^2 = e_2^2  = e_3^2 =  1 
\ee
but with each element anticommuting, that is $ e_j e_k = - e_k e_j $, for $ j \ne k $.  We then find that the composite algebraic element, the trivector $ \iGA = e_1 e_2 e_3 $ squares to minus one, that is, $ \iGA^2 = (e_1 e_2 e_3)^2 = e_1 e_2 e_3 e_1 e_2 e_3 = - e_1 e_1 e_2 e_2 = -1 $, assuming an associative algebra, and as it is found to commute with all other elements it can be used interchangeably with the unit imaginary $ \rm{i} = \sqrt{-1} $. 

A general Clifford multivector for three-space can be written by combining all available algebraic elements
\be \label{generalMultivector}
a + x_1 e_1 + x_2 e_2 + x_3 e_3 + \iGA ( t_1 e_1 + t_2 e_2 + t_3 e_3) + \iGA b ,
\ee
where $ a $, $ b $, $ x_k $ and $ t_k $ are real scalars, and $ \iGA $ is the trivector.  We thus have eight degrees of freedom present, in which we use $ \mathbf{x} = x_1 e_1 + x_2 e_2 + x_3 e_3 $ to represent a Cartesian-type vector, and the bivector $ \iGA ( t_1 e_1 + t_2 e_2 + t_3 e_3) = t_1 e_2 e_3 + t_2 e_3 e_1 + t_3 e_1 e_2 $ to represent time, forming a time vector $ \mathbf{t} = t_1 e_1 + t_2 e_2 + t_3 e_3 $. 
That is $ \bigwedge \Re^3  $ is the exterior algebra of $ \Re^3 $ which produces the space of multivectors $ \Re \oplus \Re^3 \oplus \bigwedge^2 \Re^3 \oplus \bigwedge^3 \Re^3 $, an eight-dimensional real vector space denoted by $ Cl_{3,0}(\Re) $.   

\section*{\large{Clifford multivector spacetime}}  \label{results}

Spacetime events can be described with the multivector of three dimensions \cite{chappell2011revisiting} 
\be \label{spacetimeEventv2}
X = \mathbf{x} + \iGA c \mathbf{t} ,
\ee
with $ \iGA = e_1 e_2 e_3 $ the trivector.  In order to retain the correct spacetime distance we apply a constraint of orthogonality between the space and the time three-vectors, through stipulating $ \mathbf{x} \cdot \mathbf{t} = 0 $. Now, squaring the multivector we find
\bea \label{fullXsquaredv2perp}
X^2 & = & (\mathbf{x} + \iGA c \mathbf{t})(\mathbf{x} + \iGA c \mathbf{t}) \\ \nonumber
& = & \mathbf{x}^2 - c^2 \mathbf{t}^2  + 2 \iGA c \mathbf{x} \cdot \mathbf{t} = \mathbf{x}^2 - c^2 \mathbf{t}^2 \nonumber
\eea
using the fact that $ \mathbf{x} \mathbf{t} +  \mathbf{t} \mathbf{x} =  2 \mathbf{x} \cdot \mathbf{t} = 0 $, referring to Eq.~(\ref{dotwedgeform}) in the Appendix, thus producing the correct spacetime distance in three dimensions. 

\subsection{Lorentz invariance}

The Lorentz transformations describe the transformations for observations between inertial systems in relative motion. It was found that the homogeneous Lorentz group was defined through the operators, described by a boost velocity $ \mathbf{v} $ and rotation $ \theta $ about an $  \hat{\mathbf{w}} $ axis, given by 
\be \label{HomLorentzGroup}
L = \rme^{\phi \hat{\mathbf{v}}} \rme^{ i  \hat{\mathbf{w}} \theta } ,
\ee
Defining the dagger operation $ L^{\dagger} = \rme^{ -i  \hat{\mathbf{w}} \theta }  \rme^{-\phi \hat{\mathbf{v}}} $, we find $ L L^{\dagger} = 1 $ with the transformed coordinates 
\be \label{coordinateBoosts}
X' = \rme^{ -i  \hat{\mathbf{w}} \theta /2} \rme^{- \hat{\mathbf{v}} \phi /2 } X \rme^{  \hat{\mathbf{v}} \phi /2 } \rme^{ i  \hat{\mathbf{w}} \theta/2 }.
\ee
The invariance of the spacetime distance is then easily seen from
\be \label{expMultivectorInvariant}
X'^2  = L X L^{\dagger} L X L^{\dagger}  =  L X^2 L^{\dagger}  = X^2 ,
\ee
using associativity and the fact that $ X^2 $ is a scalar as shown in Eq.~(\ref{fullXsquaredv2perp}), and so is unaffected by the transformation $ L $. 

\section*{\large{Galaxy rotation curves}} \label{darkmatter} 

Exact solutions are known for Einstein's equations of general relativity, in several special cases, such as a metric for a static field given by
\be \label{weakfieldmetricfull}
d S^2 =  \left ( 1 - \frac{2 \phi}{c^2} \right ) \left (dx^2 + d y^2 + d z^2 \right ) -\left ( 1 + \frac{2 \phi}{c^2} \right ) c^2 d t^2 .
\ee
This metric when compared with our model for spacetime in Eq.~(\ref{spacetimeEventv2}) leads us to define a spacetime event in curved space as
\be \label{dS5D}
d S = \left ( 1 - \frac{2 \phi}{c^2} \right )^{1/2}  d\mathbf{x} + \iGA  \left ( 1 + \frac{2 \phi}{c^2} \right )^{1/2}  c d \mathbf{t} .
\ee
Squaring $ d S $ we find  
\bea \label{spacetime3Dsquared} \nonumber
 d S^2 & = &  \left ( 1 - \frac{2 \phi}{c^2} \right ) d\mathbf{x}^2 - \left ( 1 + \frac{2 \phi}{c^2} \right )  c^2 d \mathbf{t}^2 +  \iGA  c \left ( 1 - \frac{4 \phi^2}{c^4} \right )^{1/2}  \left ( d\mathbf{x}  d \mathbf{t} + d\mathbf{t}  d \mathbf{x} \right ) \\ 
 & = & \left ( 1 - \frac{2 \phi}{c^2} \right ) d\mathbf{x}^2 - \left ( 1 + \frac{2 \phi}{c^2} \right )  c^2 d \mathbf{t}^2 
\eea
using $ d\mathbf{x}  d \mathbf{t} + d\mathbf{t}  d \mathbf{x} = 0 $ producing the required result in Eq.~(\ref{weakfieldmetricfull}), except that we have now defined time as a three-vector.  For galaxy rotation curves we are in the non-relativistic regime with $ v \ll c $, and so we can neglect the spatial corrections to the metric, producing $  d S^2 = d\mathbf{x}^2 - \left ( 1 + \frac{2 \phi}{c^2} \right )  c^2 d \mathbf{t}^2 $, where $ \phi $ now becomes identifiable with the Newtonian potential $ \phi = - \frac{G M}{r} $.

The required orthogonality of the space $ \mathbf{x} $ and time vectors $ \mathbf{t} $, given a spatial direction $ \mathbf{x} $, effectively reduces the degrees of freedom in the time vector by one, so writing Eq.~(\ref{spacetime3Dsquared}) in spherical coordinates with an additional time dimension $ t' $ we find
\be
d S^2 =  d r^2 +  r^2 \left (d \theta^2 + \sin^2 \theta \, d \rho^2  \right ) -\left ( 1 + \frac{2 \phi}{c^2} \right ) c^2 d t^2 -\left ( 1 + \frac{2 \phi}{c^2} \right ) c^2 d t'^2 ,
\ee
valid for a static weak field in the non-relativistic limit.

Now we can identify the first time component $ t $ as the time taken for light at a speed $ c $ to cross a given distance of space, where we utilize Einstein's postulate of the invariance of $ c $ with respect to inertial observers.  We know that light received at a time, say  $ t_r $ from a source at a distance $ \mathbf{r} $, was emitted at an earlier time $ t_e = t_r - \frac{r}{c} $.  However if clocks ran at slower rates at earlier times then we would also need to add an extra contribution for the time difference between these points.  If we look for observational evidence for this idea we can immediately find it within the red-shift data, typically interpreted as evidence for the Hubble expansion of the universe.
The Hubble expansion rate is typically calculated to be approximately $ H_0 = 70~{\rm{km/s/Mparsec}} $.  The units indicate a velocity difference between an observer and object increasing with separation distance, that is, an apparent acceleration. So dividing the distance by the speed of light we find the Hubble expansion in S.I. units of acceleration as $ a_0 = c H_0 = 6.9 \times 10^{-10}~{\rm{m/s^2}} $. The early universe is believed to have an early inflationary phase, but subsequent to this we can assume a uniform acceleration $ a_0 = \frac{d v}{ d t} $ giving $ d t = \frac{1}{a_0} d v $ and hence we find that $ c^2 d t'^2 = c^2 \left( \frac{1}{a_0^2} dv^2 \right ) = dv^2/H_0^2 $, where $ v $ represents the Hubble expansion velocity.  This then gives the metric
\be
d S^2 =  d r^2 +  r^2 \left (d \theta^2 + \sin^2 \theta \, d \rho^2  \right ) -\left ( 1 + \frac{2 \phi}{c^2} \right ) c^2 d t^2 -\left ( 1 + \frac{2 \phi}{c^2} \right ) \frac{d v^2 }{H_0^2},
\ee
where the additional time contribution to the metric is expressed by the velocity difference divided by the Hubble expansion rate $ H_0 $.
For a spherically symmetric galaxy we have $ d \theta = d \rho = 0 $, and for convenience we write $ \tau = 1/H_0 \approx 13.8 \times 10^9~{\rm{years}}$, which is the approximate age of the universe assuming a constant Hubble expansion, thus we find
\be \label{finalspacetimemetric}
d S^2 =  d r^2  -\left ( 1 + \frac{2 \phi}{c^2} \right ) c^2 d t^2 -\left ( 1 + \frac{2 \phi}{c^2} \right ) \tau^2 d v^2.
\ee
If we define the present time as $ t = 0 $, with a spacetime distance $ d S^2 = 0 $ to represent photons, we have  $ d r^2  = \left ( 1 + \frac{2 \phi}{c^2} \right ) \tau^2 d v^2 $, giving $ \frac{d r}{d v}  = \tau \left ( 1 + \frac{2 \phi}{c^2} \right )^{1/2} \approx \tau (1 + \phi/c^2 ) \approx \tau $, and so $ \frac{d r}{d v} = \tau $ or $ r = v \tau = v/H_0 $, indicating a steadily expanding universe at the Hubble rate as required in the weak field limit.  

Now that we have obtained the five-dimensional metric in Eq.~(\ref{finalspacetimemetric}) we have arrived at a metric found by a previous author for an expanding spacetime \cite{carmelli2002}, where the extra dimension is interpreted as a velocity dimension, though subject to Lorentz transformations similar to those for time coordinates.  However with our approach, beginning from Eq.~(\ref{spacetimeEventv2}), it is clear from the outset we are dealing with a time-like coordinate and thus subject to the same Lorentz transformations, as shown in Eq.~(\ref{coordinateBoosts}).  

The equations of motion that determine the geodesic trajectory of a spherical test particle can be written
\be \label{equationsmotion}
\frac{d^2 x^{\mu} }{d s^2} + \Gamma^{\mu}_{\alpha \beta} \frac{d x^{\alpha} }{d s} \frac{d x^{\beta} }{d s},
\ee
where $ \alpha,\beta,\mu \in \{0,1,2,3,4 \} $, applicable in five dimensions, with index four referring to the second time dimension.
Noting that our metric in Eq.~(\ref{finalspacetimemetric}) is diagonal, thus we can calculate the Christoffel symbols from the standard formula $  \Gamma^{\alpha}_{\mu \nu}  =  \frac{1}{2} g^{\alpha \beta} \left ( g_{\beta \mu ,\nu} + g_{\beta \nu ,\mu} - g_{\mu \nu ,\beta} \right ) $ giving 
\bea \label{christoffelweakfield}
\Gamma^{0}_{0 0} & = & \frac{1}{2} g^{0 \beta} \left ( g_{\beta 0 ,0} + g_{\beta 0 ,0} - g_{0 0 ,\beta} \right ) = \frac{1}{2} g^{0 0}  g_{0 0 ,0} = \frac{1}{2} \frac{1}{(1+ 2\phi/c^2)} (-2 \phi/c^2 )_{,0} \\ \nonumber
\Gamma^{j}_{0 0} & = & \frac{1}{2} g^{j \beta} \left ( g_{\beta 0 ,0} + g_{\beta 0 ,0} - g_{0 0 ,\beta} \right ) = -\frac{1}{2}  \delta^{j k} g_{0 0 , k} = -\frac{1}{2} (-2 \phi /c^2)_{, j} \, ,   \nonumber
\eea
where the use of the symbols $ j, k $ in superscript or subscript, denote the spatial coordinates, that is $ j, k \in \{1,2,3 \}$.
Writing Eq.~(\ref{christoffelweakfield}) to first order we find
\be \label{christoffelsymbols}
\Gamma^{0}_{0 0} = \frac{\phi_{,0}}{c^2}  + O(\phi^2 ) \, , \,\,  \Gamma^{j}_{0 0}  =  \frac{ \phi_{,j}}{c^2} + O(\phi^2 ) . 
\ee

We find from Eq.~(\ref{equationsmotion}) the equations of motion \cite{carmelli2002}
\bea
\frac{d^2 x^j}{d t^2 } & = & - \frac{\partial \phi}{\partial j} \,\, , \, \frac{d^2 v}{d t^2 } = 0 \\ \nonumber
\frac{d^2 x^j}{d v^2 } & = & - \frac{\partial \phi}{\partial j} \,\, , \, \frac{d^2 t}{d v^2 } = 0 . \nonumber
\eea
Integrating $ \frac{d^2 v}{d t^2 } = 0 $ or $ \frac{d^2 t}{d v^2 } = 0 $ gives $ v = a_0 t $, where $ a_0 $ is a constant and can be taken equal to $ a_0 = c/\tau = c H_0 $.  This acceleration is in the same direction as the cosmic expansion $ v $ and hence will be outward if we assume circular orbits.
Now looking at $ \frac{d^2 x^j}{d t^2 } = - \frac{\partial \phi}{\partial j} $ we find the first integral $ \left ( \frac{d s}{d t} \right )^2 = \frac{G M }{r} $, as expected for a Newtonian potential in the weak field limit.  

If we now include this outward acceleration for a star orbiting a galactic gravitational potential we then have for the effective potential
\be \label{effectivePotential}
V_e = - \frac{G M m}{r} + \frac{L^2}{2 m r^2} + m a_0 r .
\ee
A circular orbit is obtained at the minimum of the effective potential that can be found by equating the differential $ d V_e /d r $ to zero, that is $\frac{d V_e }{d r} =  \frac{G M m}{r^2} - \frac{L^2}{ m r^3} + m a_0 = 0 $.
For a circular orbital velocity $ v_c $, we have $ L = m v_c r $ and multiplying through by $ r/m $ we find 
\be \label{orbitalvelocitessquared}
 v_c^2 = \frac{G M}{r}  + a_0 r ,
\ee
where we can see that the term $ a_0 r $ now modifies the standard Newtonian result flattening the rotation curves in agreement with observational data on galaxies.  

\section*{\large{Conclusion}} \label{conclusion} 

Using the Clifford multivector, given in Eq.~(\ref{spacetimeEventv2}), and incorporating the Hubble expansion, we provide a alternate explanation for galaxy rotation curves.

The extra time dimension is implemented into the metric in Eq.~(\ref{dS5D}) and produces the spacetime distance in Eq.~(\ref{spacetime3Dsquared}), which when identified with the Hubble expansion modifies the geodesic paths and provides an explanation for galaxy rotation curves and hence dark matter.
The physical justification for this extra time component added into the metric in Eq.~(\ref{spacetime3Dsquared}) is the redshift data, which can be interpreted as implying a time rate differential between different locations in the universe.  Implementing this extra time differential at the scale of individual galaxies produces a modified rotation curve as show in Fig~\ref{VelocityRadius} that has a flattened rotation curve in qualitive agreement with observational data.  It has also been shown Carmeli \cite{carmelli2002} that this metric also provides an alternate explanation to dark energy.

While a metric of the form in Eq.~(\ref{dS5D}) had been developed previously \cite{carmelli2002,Oliveira2006}, in their case the sign of the metric and the required coordinate transforms needed to be deduced separately, whereas with our approach it is clear from the outset we are dealing with a time like coordinate and thus subject to the conventional Lorentz transformations for time.  Also in their approach the metric is interpreted as forming a five-dimensional space, with three space, one time and one velocity dimension, whereas with our approach we remain within a three dimensional real space, with time identified with the bivectors of this space and thus giving the metric a more fundamental basis using just space and time components.  We can then view spacetime described by Eq.~(\ref{spacetimeEventv2}) with the space components representing the three translational freedoms of physical space and time representing the three rotational freedoms of physical space.

We also show in Appendix B how the metric in Eq.~(\ref{spacetimeEventv2}) can be used as mathematical model for {\it remote viewing} without recourse to a complexified Minkowski space for example.

Clifford's geometric algebra is a general mathematical formalism, which can be applied to many areas of physics, and because it can represent complex-like numbers using the trivector, and quaternion-like numbers through the even subalgebra, it can describe many physical phenomena over a strictly real three-dimensional space with obvious advantages in interpretation.

\appendix

\section{Geometric product} \label{GeoProduct}

Given two vectors $ \mathbf{u} = u_1 e_1 + u_2 e_2 + u_3 e_3 $ and $ \mathbf{v} = v_1 e_1 + v_2 e_2 + v_3 e_3 $, using the distributive law for multiplication over addition, as assumed for an algebraic field, we find their product
\bea \label{VectorProductExpand2DInitial}
\textbf{u} \textbf{v} & = & (u_1 e_1 + u_2 e_2 + u_3 e_3)( v_1 e_1 + v_2 e_2 + v_3 e_3) \\ \nonumber
& = & u_1 v_1 + u_2 v_2 + u_3 v_3 \\ \nonumber
& + & \iGA ( (u_2 v_3 - u_3 v_2 ) e_1+ (u_3 v_1 - u_1 v_3 ) e_2 + (u_1 v_2 - u_2 v_1 ) e_3 ) ,
\eea
where we have used the elementary properties of $ e_1,e_2,e_3 $ defined in Eq.~(\ref{orthonormality}).  We recognize $ u_1 v_1 + u_2 v_2 + u_3 v_3 $ as the conventional dot product and $ (u_2 v_3 - u_3 v_2 ) e_1+ (u_3 v_1 - u_1 v_3 ) e_2 + (u_1 v_2 - u_2 v_1 ) e_3  $ as the cross product, so that we can write
\be \label{VectorProductExpand2D}
\textbf{u} \textbf{v} = \textbf{u} \cdot \textbf{v}  + \iGA \textbf{u} \times \textbf{v} = \textbf{u} \cdot \textbf{v}  + \textbf{u} \wedge \textbf{v},
\ee
with the identity $ \iGA \textbf{u} \times \textbf{v} = \textbf{u} \wedge \textbf{v} $.
For $ \hat{\mathbf{u}} $ and $  \hat{\mathbf{v}} $ unit vectors, we have  $ \hat{\textbf{u}} \cdot \hat{\textbf{v}} = \cos \theta $ and $ \hat{\textbf{u}} \wedge \hat{\textbf{v}} = \iGA \hat{\textbf{w}} \sin \theta $, where $ \hat{\textbf{w}} $ is the orthogonal vector to $ \hat{\mathbf{u}} $ and $  \hat{\mathbf{v}} $, giving $ \hat{\textbf{u}} \hat{\textbf{v}} = \cos \theta + \iGA \hat{\textbf{w}} \sin \theta = \rme^{\iGA \hat{\textbf{w}}} $, where $ \theta $ is the angle between the two vectors.  

Using the commutivity of the dot product and the anticommutivity of the cross product we can rearrange Eq.~(\ref{VectorProductExpand2D}) to write
\be \label{dotwedgeform}
\textbf{u} \cdot \textbf{v} = \frac{1}{2} ( \textbf{u} \textbf{v}  + \textbf{v} \textbf{u} ) \,\,\, {\rm{and}} \,\,\, \textbf{u} \wedge \textbf{v} =  \frac{1}{2} ( \textbf{u} \textbf{v}  - \textbf{v} \textbf{u} ). 
\ee
We can see from Eq.~(\ref{VectorProductExpand2DInitial}), that the square of a vector $ \mathbf{v}^2 = \mathbf{v} \cdot \mathbf{v} = v_1^2 + v_2^2 + v_3^2 $, becomes a scalar quantity.
Hence the Pythagorean length of a vector is simply $ |\mathbf{v}| = \sqrt{\mathbf{v}^2} $, and so we can find the inverse vector 
\be
\mathbf{v}^{-1} =  \frac{\mathbf{v}}{\mathbf{v}^2}.
\ee

\section*{\large{Gravity in general relativity}} 

For a weak field $ \phi $ with a single time dimension we can write
\be
d S^2 =  \left ( 1 - \frac{2 \phi}{c^2} \right ) \left (dx^2 + d y^2 + d z^2 \right ) -\left ( 1 + \frac{2 \phi}{c^2} \right ) c^2 d t^2 ,
\ee
where the weak field limit implies $ 2 \phi \ll c^2 $.  This is a reasonable approximation in many cases as the exterior field of the sun, for example, satisfies $ 2 \phi < 5 \times 10^{-6} c^2 $.
Now, consider the metric $ d S^2 = A d r^2 - B c^2 d t^2 = dt^2 ( A v^2 - B c^2 ) $, where $ A \approx B \approx 1 $, then for a slow moving satellite we have $ v \ll c $ and so we can neglect the spatial distortion of the metric in this case.  Therefore we can write in the first approximation to the weak field case with low velocities
\be \label{weakfieldmetric}
d S^2 =   \left (dx^2 + d y^2 + d z^2 \right ) -\left ( 1 + \frac{2 \phi}{c^2} \right ) c^2 d t^2 ,
\ee
where $ \phi $ can be identified with the Newtonian potential $ \phi = -\frac{G M}{r} $.  To confirm this we find the geodesic distance between two event $ P_1 $ and $ P_2 $ at times $ t_1 $ and $ t_2 $ respectively
\be
\int_{P_1}^{P_2} dS = \int_{t_1}^{t_2} \frac{dS}{dt} dt = c \int_{t_1}^{t_2} \left ( 1 + \frac{2 \phi}{c^2} - \frac{v^2}{c^2} \right )^{1/2} dt 
\ee
and so to first order we find
\be
\int_{P_1}^{P_2} dS = c \int_{t_1}^{t_2} \left ( 1 + \frac{ \phi}{c^2} - \frac{v^2}{2 c^2} \right ) dt = c (t_2 - t_1 ) -\frac{1}{c} \int_{t_1}^{t_2} \left ( \frac{1}{2} v^2 - \phi  \right ) dt .
\ee
So the condition that $ \int dS $ be maximal is equivalent to $ \int_{t_1}^{t_2} \left ( \frac{1}{2} v^2 - \phi  \right ) dt $ being minimal, which is Hamiltons' principal for a particle in a gravitational potential $ \phi $.
So, beginning with the variation 
\be
\delta \int d S  = \delta \int \left | g_{\mu \nu} \frac{d x^{\mu}}{d s } \frac{d x^{\nu}}{d s} \right |^{\frac{1}{2}} d s  = \delta \int \left | g_{\mu \nu} \dot{x}^{\mu} \dot{x}^{\nu} \right |^{\frac{1}{2}} d s  =  \delta \int L d s  ,
\ee
where $ L = \left | g_{\mu \nu} \dot{x}^{\mu} \dot{x}^{\nu} \right |^{\frac{1}{2}} $ is the Lagrangian.  The Euler-Lagrange equations $ \frac{d}{d s} \left ( \frac{\partial L}{\partial \dot{x}^{\mu}} \right ) - \frac{\partial L}{\partial x^{\mu}} = 0 $ will now give us the equations of motion.  In order to remove the square root sign over the metric we define a new Lagrangian $ {\cal{L}} = L^2 = \frac{d S^2}{d s^2} = g_{\mu \nu} \dot{x}^{\mu} \dot{x}^{\nu} $ that can be shown to produce the same set of Euler-Lagrange equations as for $ L $, and so we arrive at the equations of motion that determine the geodesic trajectory of a spherical test particle 
\be \label{equationsmotionappendix}
\frac{d^2 x^{\mu} }{d s^2} + \Gamma^{\mu}_{\alpha \beta} \frac{d x^{\alpha} }{d s} \frac{d x^{\beta} }{d s},
\ee
where $ \alpha,\beta,\mu \in \{0,1,2,3 \} $.
However we have the four momentum $ p = m \frac{d x}{d \tau} $ and $ \tau $ is an affine parameter of the curve.
Hence we have an alternate form for the geodesic equations as
\be
p^{\alpha} p^{\mu}_{,\alpha} + \Gamma^{\mu}_{\alpha \beta}  p^{\alpha}  p^{\beta} = 0,
\ee
where the notation $  p^{\mu}_{,\alpha} $ indicates a partial differentiation of $ p^{\mu} $ with respect to coordinate $ \alpha $. If we firstly take the zeroeth order equation in momentum (representing energy) we have
\be
p^{\alpha} p^{0}_{,\alpha} + \Gamma^{0}_{\alpha \beta}  p^{\alpha}  p^{\beta} = 0 .
\ee
For non-relativistic particles we have $ p^{0} \ll p^{j} $ and $ p^{\alpha} \partial_{\alpha} = m U^{\alpha} \partial_{\alpha} \approx  m U^{0} \partial_{0} \approx m c \frac{d}{d \tau} $, so we have approximately
\be
m \frac{d p^{0}}{d \tau} + \Gamma^{0}_{0 0} ( p^{0})^2 = 0 . 
\ee
But $ (p^{0})^2 = (p^j)^2 + m^2 c^2 \approx m^2 c^2 $ and $ t \approx \tau $ for low momentum, so that we have
\be
\frac{d p^{0}}{d \tau} = - m \frac{\partial \phi}{\partial \tau },
\ee
which states that the energy is conserved if the field does not depend on time.

The spatial components of the geodesic equations give
\be
p^{\alpha} p^{j}_{,\alpha} + \Gamma^{j}_{\alpha \beta}  p^{\alpha}  p^{\beta} = 0 
\ee
that for non-relativistic particles becomes approximately $ m \frac{d p^{j}}{d \tau}  + \Gamma^{j}_{0 0 }  (p^{0 })^2 = 0 $.
Hence we have
\be
\frac{d p_{j}}{d \tau} = - m \phi_{,j} = - m \nabla \phi,
\ee
which is the conventional expression for the gravitational force in Newtonian theory.  Hence we see how the metric in Eq.~(\ref{weakfieldmetric}) produces the results of Newtonian theory in the weak field limit.  We can now extend this metric to include a second time dimension as indicated by the spacetime metric in Eq.~(\ref{spacetimeEventv2}) to see how it alters geodesic paths for stars around a galactic gravitational centre.

\section*{Expanded metric} 

If we expand the weak field metric in Eq.~(\ref{weakfieldmetric}) to include a second time dimension $ t' $ and writing Eq.~(\ref{spacetime3Dsquared}) in spherical coordinates we find
\be
d S^2 =  d r^2 +  r^2 \left (d \theta^2 + \sin^2 \theta \, d \rho^2  \right ) -\left ( 1 + \frac{2 \phi}{c^2} \right ) c^2 d t^2 -\left ( 1 + \frac{2 \phi}{c^2} \right ) c^2 d t'^2 .
\ee

The redshift phenomena, presumed to be due to the expansion of the universe, is strictly speaking not a Doppler shift phenomena, but it is conventionally assumed that the redshift is caused by the expansion of space itself not the relative motion of galaxies within this space. That is, the expanding space of the universe creates apparent velocity differentials between points throughout this space as modeled by Eq.~(\ref{finalspacetimemetric}). 
 However alternate interpretations of this metric are possible.  For example, if it is discovered that clocks (as measured by atomic orbitals for example) ran at slower rates earlier in the universe's history, then  we could obtain a static universe with the redshift caused by observing galaxies in an earlier period in history due to the time delay caused by the finite propagation speed of light.

We assume Einstein's field equations $ R_{\mu \nu} -\frac{1}{2} g_{\mu \nu} R = \kappa T_{\mu \nu} $ expand unchanged to five dimensions with $ \mu,\nu \in \{0,1,2,3,4\} $ , with an added component (index four), describing the second time dimension.
The equations of motion from Eq.~(\ref{equationsmotion}) in the weak field limit also expanded to five dimensions, using the same approximation used earlier, produce
\bea
\frac{d^2 x^{\mu}}{d t^2 }+ \Gamma^{\mu}_{\alpha \beta} \frac{d x^{\alpha} }{d t} \frac{d x^{\beta} }{d t} & = & 0 \\ \nonumber
\frac{d^2 x^{\mu}}{d v^2 }+ \Gamma^{\mu}_{\alpha \beta} \frac{d x^{\alpha} }{d v} \frac{d x^{\beta} }{d v} & = & 0 .\nonumber
\eea
Non-relativistically we only need to retain $ \Gamma^{\mu}_{0 0} $ in the first approximation because $ v \ll c$ and so $ d x^j/d t \ll d x^0/dt $ and so
\bea \label{eqnmotion3d}
\frac{d^2 x^{\mu}}{d t^2 } & \approx & - \Gamma^{\mu}_{00} \left (\frac{d x^{0} }{d t} \right )^2 = - \Gamma^{\mu}_{00} c^2 = - \frac{\partial \phi}{\partial \mu} \\ \nonumber
\frac{d^2 x^{\mu}}{d v^2 } & = &  - \Gamma^{\mu}_{0 0} \left( \frac{d x^{0} }{d v} \right )^2 , \nonumber
\eea
using the evaluation of the Christoffel symbols in Eq.~(\ref{christoffelsymbols}).
The first equation in Eq.~(\ref{eqnmotion3d}) splits into
\bea
\frac{d^2 x^j}{d t^2 } & = & - \frac{\partial \phi}{\partial j} \,\, , \, \frac{d^2 v}{d t^2 } = 0 \\ \nonumber
\frac{d^2 x^j}{d v^2 } & = & - \frac{\partial \phi}{\partial j} \,\, , \, \frac{d^2 t}{d v^2 } = 0 . \nonumber
\eea
Integrating $ \frac{d^2 v}{d t^2 } = 0 $ or $ \frac{d^2 t}{d v^2 } = 0 $ gives $ v = a_0 t $, where $ a_0 $ is a constant and can be taken equal to $ a_0 = c/\tau = c H_0 $.  
Now looking at $ \frac{d^2 x^j}{d v^2 } = - \frac{\partial \phi}{\partial j} $ we find the first integral $ \left ( \frac{d s}{d v} \right )^2 = \frac{k M }{r} $ in analogy to the Newtonian equation $ \left ( \frac{d s}{d t} \right )^2 = \frac{G M }{r} $.  Comparing these we find $ \frac{d s}{d v} = \frac{\tau}{c} \frac{d s}{d t} $ or $ \frac{d v}{d t} = \frac{c}{\tau} $, that is $ a_0 = c/\tau $ as obtained previously.

\begin{figure}[htb]
\begin{center}
\includegraphics[width=5.2in]{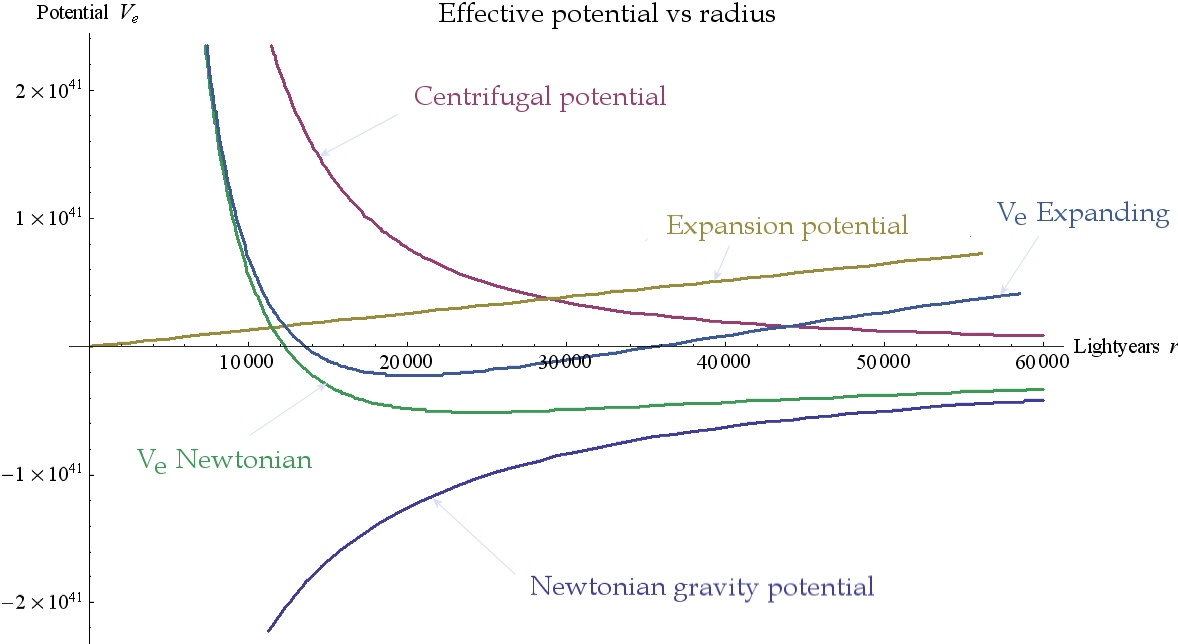}
\end{center}

\caption{The effective potential for a star orbiting a galaxy approximately the size of the Milky way galaxy, showing a minimum in the effective potential at around 20,000 light years (the sun lies at approximately 26,000 light years).  Including the cosmic expansion into the effective potential we see a slight lifting of the effective potential above the classical Newtonian potential and a movement to a tighter orbit. \label{EffectivePotential}}
\end{figure}

If we include the outward acceleration for a star orbiting a galactic gravitational potential we then have the effective potential given in Eq.~(\ref{effectivePotential}) and graphed in Fig.~\ref{EffectivePotential}.

\begin{figure}[htb]

\begin{center}
\includegraphics[width=4.6in]{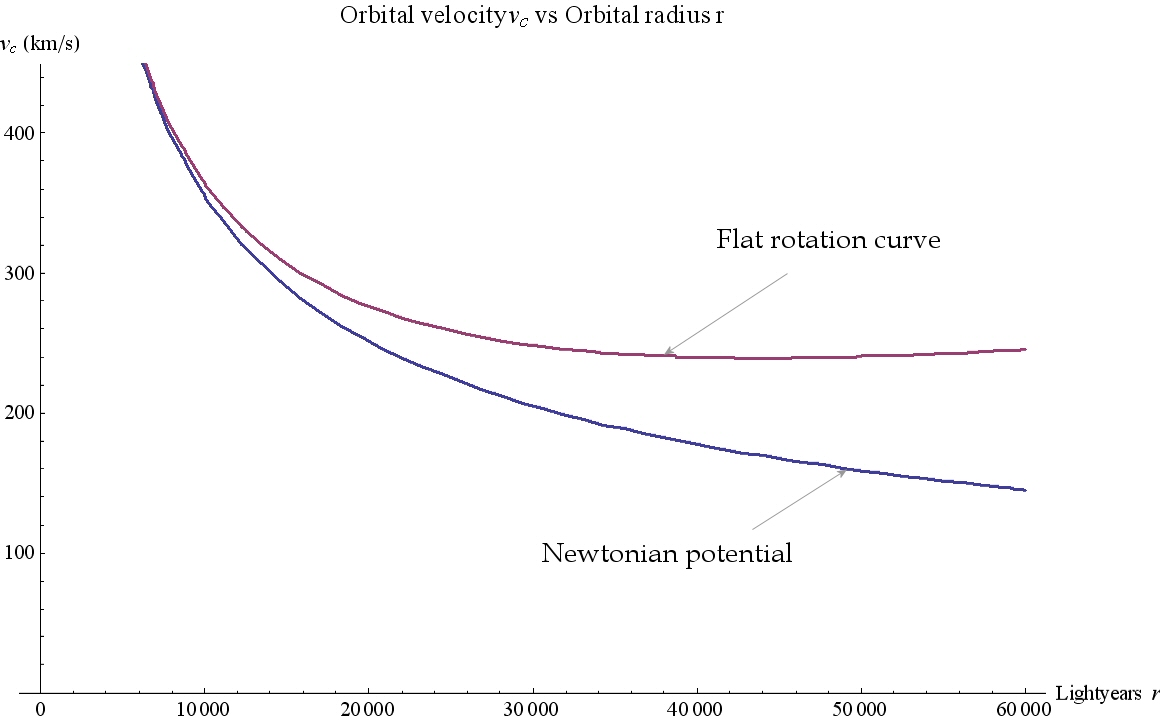}
\end{center}

\caption{For a star orbiting a galaxy the size of the Milky Way galaxy we see how the velocity falls as $ \frac{G M}{r} $ for a Newtonian potential(blue), but including the expansion potential from the second time dimension we find a flat rotation curve(red) as observed. The sun has an orbital speed of approximately 240 km/s at a radius of 26,000 light years, lying fairly close to this curve. We are assuming that all the mass lies inside the orbital radius and so this is only an approximate diagram but shows the effect of the expansion in flattening the rotation curve. \label{VelocityRadius}}
\end{figure}

Squaring Eq.~(\ref{orbitalvelocitessquared}) we find
\be \label{finalvcrelationship}
v_c^4 = \left( \frac{G M}{r} \right )^2 + 2 G M a_0 + a_0^2 r^2 ,
\ee
as found previously \cite{carmelli2002}.
The first term is the Newtonian contribution, the second term is a new contribution from the outward acceleration and the third term is relatively small when considering orbits within galaxies.  If we assume the luminosity $ L $ of a galaxy  is roughly proportional to its total mass $ M $ then Eq.~(\ref{finalvcrelationship}) implies the relationship $ v_c^4 \propto L $, which is the Tully-Fisher law for galaxy rotation curves.  If we graph the orbital velocities as a function of the distance from the galactic centre as given by Eq.~(\ref{orbitalvelocitessquared}) we find nearly flat rotation curves, as shown in Fig.~\ref{VelocityRadius}. Hence the Clifford multivector description of spacetime, shown in Eq.~(\ref{spacetimeEventv2}) naturally produces the correct galaxy rotation curves without the need for a dark matter hypothesis.

\section{\large{Application to remote viewing}} 

Consider two events described by Eq.~(\ref{spacetimeEventv2}), by writing 
\be \label{spacetimeEventv2para1}
X_1 = \mathbf{x}_1 + \iGA \mathbf{t}_1 , X_2 = \mathbf{x}_2 + \iGA \mathbf{t}_2 ,
\ee
with time now measured in meters in order to remove the extra constant of $ c $.
So that we have the separation
\be \label{spacetimeEventv2parapsychology}
\Delta S = X_2 - X_1 = \mathbf{x}_2 - \mathbf{x}_1 + \iGA \left ( \mathbf{t}_2 - \mathbf{t}_1 \right ) .
\ee
Causality is specified by remaining on the light cone, requiring $ (\Delta S)^2 = 0 $.  That is
\be \label{spacetimeEventv2parapsychologyv2}
(\Delta S )^2 = \left (\mathbf{x}_2 - \mathbf{x}_1 \right )^2 - \left ( \mathbf{t}_2 - \mathbf{t}_1 \right )^2 = 0 .
\ee

If we attempt to remote view some distant location, without loss of generality, we align the axes such that the location lies in the $ e_1 $ direction, and therefore the time vector being orthogonal will consist of the two components $ s e_2 + t e_3 $.
We have $ \left (\mathbf{x}_1 - \mathbf{x}_2 \right ) = d e_1 $, the distance and direction to the location, and if we assume that the light travels in the $ i e_3 t = e_1 e_2 t $ time direction, and we require $ t = 0 $ in order for the remote viewing to be instantaneous,  we are then left with the causal requirement
\be \label{causalremoteviewing}
(\Delta S )^2 = d^2 - s^2 = 0.
\ee
However this can be satisfied if somehow an instrument can separate out the two time components, and utilize the time direction $ s i e_2 $, and view through a time distance $ s = d $.
Hence the metric of spacetime described within the Clifford multivector provides a simple framework for further investigation.  Presently this remains purely an intriguing theoretically construct, given there are no known measuring instruments that can perform the required function.

\bigskip
	
%\section*{Acknowledgements} \label{ack} 

%\section*{Author contributions} \label{authorcontrib}

%Original idea by J.C., following which all authors have contributed equally to this work. 	

%\section*{Additional information} \label{additionalinfo}

%Competing financial interests: The authors declare no competing financial interests.

%Attribution-NonCommercial-ShareAlike 3.0 Unported License. To view a copy of this
%license, visit http://creativecommons.org/licenses/by-nc-sa/3.0/

%How to cite this article: Chappell, J., Iannella, N., Iqbal, A. \& Abbott, D. A new description of space and time using Clifford %multivectors. 

%\bibliographystyle{abbrvnat}
%\bibliographystyle{plos2009}
\bibliographystyle{model1a-num-names}

\bibliography{quantum}

\end{document}